%% file: main_ASILOMAR.tex
\documentclass[conference]{IEEEtran}

\usepackage{graphicx} % Required for inserting images

\usepackage{color,array,amsthm}
\usepackage{amsmath,amssymb,amsfonts}
\usepackage[utf8]{inputenc}
\usepackage{amssymb}
\usepackage[table]{xcolor}
\usepackage{graphicx}
\usepackage[nolist]{acronym}
\usepackage{color}
\usepackage{relsize}
\usepackage{threeparttable}
\usepackage[nolist]{acronym}
\usepackage{multirow}
\usepackage{framed}
\usepackage{soul}
\usepackage{algpseudocode}
\usepackage{algorithm}
\usepackage{amsmath}
\usepackage{xcolor}
\usepackage{balance}
\usepackage{comment}
\usepackage{longtable}
\usepackage{graphicx}
\usepackage{booktabs, tabularx}
\usepackage{arydshln}
\usepackage{subcaption}
\usepackage{units}

\newcommand{\nic}[1]{\textcolor{blue}{\textbf{ND:} #1}}

\begin{document}
%\bstctlcite{IEEEexample:BSTcontrol}

\title{Ziv-Zakai Bound for \\ Near-Field Localization and Sensing
}

\author{\IEEEauthorblockN{
Nicol\`o Decarli\IEEEauthorrefmark{1}\IEEEauthorrefmark{3},
Davide Dardari\IEEEauthorrefmark{2}\IEEEauthorrefmark{3}
}\\

\IEEEauthorblockA{\IEEEauthorrefmark{1} IEIIT, National Research Council (CNR), 40136 Bologna, Italy}
\IEEEauthorblockA{\IEEEauthorrefmark{2} DEI, Universit\`a di Bologna, 40136 Bologna, Italy}
\IEEEauthorblockA{\IEEEauthorrefmark{3} National Laboratory of Wireless Communications (WiLab), CNIT, 40136 Bologna, Italy}
}

\maketitle

\input{acronyms}

\input{macros}

\begin{abstract}
The increasing carrier frequencies and growing physical dimensions of antenna arrays in modern wireless systems are driving renewed interest in localization and sensing under near-field conditions. In this paper, we analyze the Ziv-Zakai Bound (ZZB) for near-field localization and sensing operated with large antenna arrays, which offers a tighter characterization of estimation accuracy compared to traditional bounds such as the Cramér-Rao Bound (CRB), especially in low signal-to-noise ratio or threshold regions. Leveraging spherical wavefront and array geometry in the signal model, we evaluate the ZZB for distance and angle estimation, investigating the dependence of the accuracy on key signal and system parameters such as array geometry, wavelength, and target position. Our analysis highlights 
the transition behavior of the ZZB and underscores the fundamental limitations and opportunities for accurate near-field sensing. \end{abstract}

%\tableofcontents

\section{Introduction}

We focus on the problem of near-field localization by adopting large antenna arrays. This scenario is increasingly relevant in modern communication, radar, and sensing systems, where the transmitter to be localized or a target lie within the near-field region of a receiving array \cite{CheEtAl:24,WanEtAl:J25}.
A standard way to assess performance is through lower bounds on the \ac{MSE} related to the estimation performance, which provide benchmarks for estimator design and reveal fundamental performance limits.
The \ac{CRB} is the most widely used tool \cite{Kay:93}. It is asymptotically tight, as the \ac{MLE}, which is often adopted in practice, approaches it at high \ac{SNR}. However, \ac{CRB} is accurate only when the estimation errors are small and confined close to the true value of the parameter. This assumption holds in the high-\ac{SNR} regime, where estimation errors are typically small perturbations induced by noise \cite{Kay:93}. 
Since, in the near-field localization problem, the received signal depends non-linearly on the unknown parameters to be estimated, such as distance and \ac{AoA}, at low or moderate \ac{SNR} the behavior of the estimator differs significantly. In fact, in this condition, sidelobes in the likelihood function induce ambiguities that cause large estimation errors, which the \ac{CRB} cannot capture.

As an alternative, the \ac{ZZB} \cite{ZivZak:69, BelTar:74, ChaZakZiv:75} is a Bayesian bound, which assumes that the parameter of interest is a random variable with a known a-priori distribution. 
The \ac{ZZB}, frequently adopted as performance benchmark for time-based distance estimation adopting wideband signals \cite{SadLiuXu:J10,GifDarWin:J22,ZhaEtAL:J23}, is known to effectively capture the so-called \textit{threshold effect}, where estimator performance undergoes a sharp transition from being dominated by noise perturbations to being dominated by global ambiguities. The correct prediction of the location of these thresholds is crucial for system design and performance evaluation.

Motivated by these considerations, this paper investigates the \ac{ZZB} as a way to characterize the performance limits of near-field localization and sensing with large antenna arrays. Unlike previous works that assumed sparse sensors \cite{ChiHaiSchDab:C09,ChiHai:C10}, here we consider a \ac{ULA} aiming at localizing a transmitter or a radar target lying within its near-field region, i.e., at a distance below the so-called \textit{Fraunhofer distance}, for which the corresponding \ac{CRB} expressions are known in closed form. Our investigation sheds light on the \ac{ZZB} in the same condition, highlighting its ability in lower bounding the actual estimation error more tightly in a wide \ac{SNR} range. This provides new insights into the design and performance evaluation of next-generation near-field array processing for localization and sensing.

\section{System Model}

\begin{figure}[t]
	\centerline{\includegraphics[width=0.99\columnwidth]{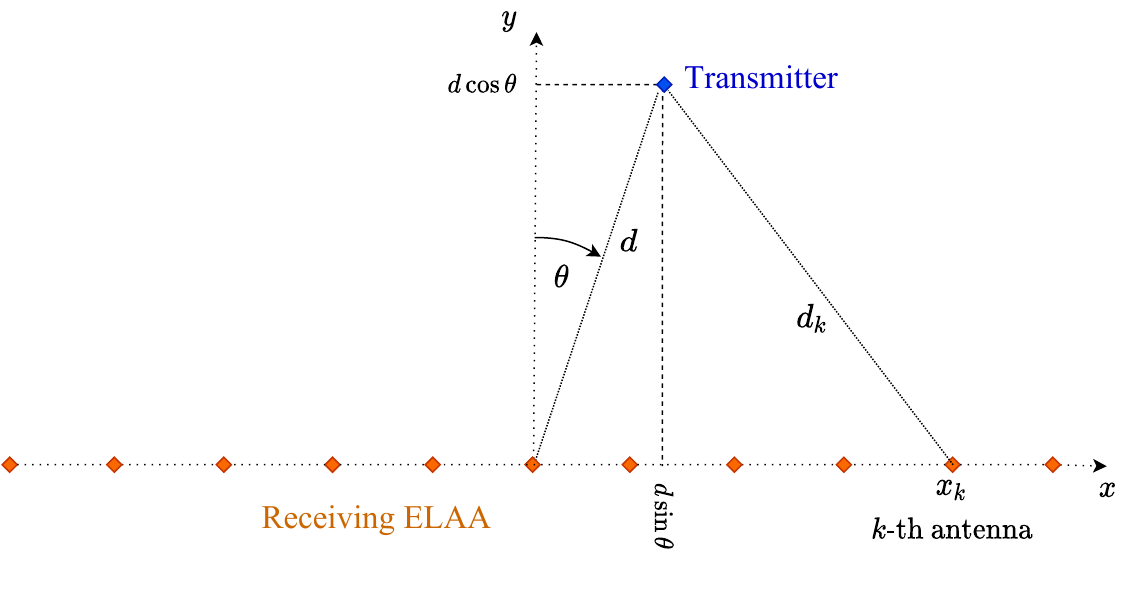}}
	\caption{Considered 2D scenario, where an ELAA is employed to localize a single antenna transmitting UE in $\boldp=[d,\theta]^{\top}$.}
	\label{fig:Scenario}
\end{figure}

We consider a single-antenna \ac{UE} that transmits a narrowband signal (it can represent a subcarrier of an \ac{OFDM} signal) at frequency $\fc$.\footnote{As shown in \cite{GioDecZanDar:C25}, the same model applies to target localization in radar, where the target reflects the transmitted signal.} The transmitter is placed in position $\tilde{\mathbf{p}} = \left [ x, y\right ]^{\top}$ which in polar coordinates corresponds to $\boldp=[d,\theta]^{\top}$ (see Fig.~\ref{fig:Scenario}). 
A \ac{BS} equipped with an \ac{ELAA} is present in the scenario. Specifically, we consider a \ac{ULA} with $K$ antennas and antenna spacing $\delta$. The \ac{ELAA} has an aperture $\Da = (K - 1)\delta$. The reference system is placed on the central element of the receiving \ac{ELAA}, which acts as reference antenna. The $k$-th element of the receiving \ac{ELAA} is located at $\tilde{\mathbf{p}}_{k} = \left [ x_k,y_k\right ]^{\top}$, with $k = -(K-1)/2, \dots , (K-1)/2$ assuming $K$ odd so that the central antenna index is 0.

The goal of the \ac{BS} is to estimate the position $\boldp$ by processing the signal received at the \ac{ELAA}. 
The signal received by the \ac{ELAA} in near-field conditions (spherical wavefront) can be written as
\begin{equation}\label{eq:RX}
\boldr=\alpha \, \bolds(\boldp) + \boldn = \alpha \, \bolds(d,\theta) + \boldn
\end{equation}
where $\alpha$ is the signal amplitude, $\bolds(d,\theta)\in \mathbb{C}^K$ is the near-field steering vector which is a function of $\boldp=[d,\theta]^{\top}$, and $\boldn\sim\mathcal{CN}(\mathbf{0},\sigma^2\mathbf{I}_K) \in \mathbb{C}^K$ denotes the \ac{AWGN}. 
Specifically, the $k$-th entry of the near-field steering vector is
\begin{equation}\label{eq:sk}
\left[\bolds(d,\theta)\right]_k=e^{-\jm \frac{2\pi}{\lambda} d_k(d,\theta)}
\end{equation}
being $\lambda=c/\fc$ the wavelength, with $c$ denoting the speed of light, and $d_k$ the distance between the \ac{UE} and the $k$-th receiving antenna which can be expressed as
\begin{align}\label{eq:dk}
d_k(d,\theta)&=\| \tilde{\boldp}-\tilde{\boldp}_k\|
= d\sqrt{1+\frac{x_k^2}{d^2}-\frac{2x_k\sin{\theta}}{d}}.
\end{align}
%
\begin{comment}
The conditional \ac{pdf} of $\boldr$ is given by
%
\begin{equation} \label{eq:pdf}
	\pX{\boldr|d,\theta}=\frac{1}{\left(\pi \sigma^2\right)^{K}} 
	\exp \left \{  -\frac{1}{\sigma^2} {\left\| \boldr - \alpha\, \bolds(d,\theta) \right \|}^2   \right\} \, .
\end{equation}
\end{comment}

By resorting to the Fresnel approximation in the near-field steering vector $\bolds(d,\theta)$, closed-form expressions for the \ac{CRB} for distance and \ac{AoA} estimation can be derived analytically, obtaining \cite{KorsoCRB:J10}
\begin{equation}\label{CRBtradRange}
\CRB^{( \hat{d} )} \!=\! \frac{6c^2d^2\left ( \delta^2\left ( K^2\!-\!4\right )\sin^2{\theta} + 15d^2\right )}{\pi^2 \!\fc^2 K \, \SNR \cos^4{\theta} \, \delta^4 \left ( K^2\!-\!4\right ) \left ( K^2 \!-\!1\right )} ,
\end{equation}
\begin{equation}\label{CRBtradAOA}
\CRB^{(\hat{\theta})} = \frac{3c^2}{2\pi^2 \fc^2 K \, \SNR \cos^2{\theta} \, \delta^2 (K^2-1)}\, 
\end{equation}
where the \ac{SNR} at the single receiving antenna is defined as $\SNR = \alpha^2/\sigma^2$.

\section{Ziv-Zakai Bound}

Consider now that the \ac{UE} may be located at a distance within the range $[\dmin, \dmax]$ and at an angle within the range $[\tmin, \tmax]$.  In the absence of any knowledge about the distribution of these parameters, it is reasonable to assume a uniform prior distribution over the specified intervals, i.e.,
\begin{equation}\label{eq:pdf_p}
%p_{d,\theta}(d,\theta)
p_{\boldp}(\boldp) = \begin{cases}
\frac{1}{T_{\mathrm{d}}T_{\mathrm{\theta}}}, & \text{if } d \in [\dmin, \dmax], \,\, \theta \in [\tmin, \tmax] \\
0, & \text{otherwise} \, 
\end{cases}
\end{equation}
where we have defined the distance span as $T_{\mathrm{d}}=\dmax-\dmin$,
and the angular span as $T_{\theta}=\tmax-\tmin$.
We notice that the \acp{CRB} in \eqref{CRBtradRange} and \eqref{CRBtradAOA} are \textit{local bounds}, i.e., they depend on the specific values of the parameters to be estimated $d$ and $\theta$, and they do not take into account the a-priori information about the parameter space.

Let us consider the joint estimation of both distance $d$ and \ac{AoA} $\theta$, i.e., of the 2-dimensional continuous vector random parameter $\boldp$ based on noisy observation $\boldr$. In such a case, the estimation error is $\boldepsilon=[\widehat{d}-d,\widehat{\theta}-\theta]^{\top}$, where $\widehat{\boldp}(\boldr)=[\widehat{d}, \widehat{\theta}]^{\top}$ is the estimated vector, and the error correlation matrix is
\begin{equation}
    \boldR_{\epsilon}=\EX{(\widehat{\boldp}(\boldr)-\boldp)(\widehat{\boldp}(\boldr)-\boldp)^{\top}}
\end{equation}
with the expectation taken over $\boldr$ and $\boldp$.
Lower bounds on the estimation \ac{MSE} of the individual components of $\boldp$ can be taken by lower bounding $\bolda^{\top}\boldR_{\epsilon}\bolda$ when $\bolda$ is a unit vector with a unit in the first (for distance) or second (for \ac{AoA}) position.
Following the derivation proposed for a vector parameter in \cite{BelSteEphVan:J97}, we get the general expression for the vector \ac{ZZB} as
\begin{align}\label{eq:ZZB2Dgen}
&\bolda^{\top} \boldR_{\epsilon} \bolda  \geq \\
 &\frac{1}{2} \!\int_0^{\infty} \!\!\!\!\max _ { \bolddelta : \bolda^{^{\top}} \bolddelta = h } \iint \left(p_{\boldp}(\boldp)\!+\!p_{\boldp}(\boldp\!+\!\bolddelta)\right) P_{\min }(\boldp, \boldp\!+\!\bolddelta) \, \dd \boldp \, h \, \dd h\, \nonumber
\end{align}
where $\bolddelta=[\Delta_{\mathrm{d}}, \Delta_\theta]^{\top}$ is a certain distance-angle displacement with respect to position $\boldp$.
The \ac{ZZB} requires the evaluation of the error probability $\Pmin{\boldp, \boldp\!+\!\bolddelta}$ corresponding to the decision problem
\begin{subequations}\label{eq:test}
\begin{align}
\Hone: & \quad \boldr=\alpha \, \bolds(\boldp) + \boldn   \, \\ 
\Htwo: & \quad  \boldr=\alpha \, \bolds(\boldp + \bolddelta) + \boldn 
\end{align}
\end{subequations}
using the optimum binary detector, i.e., the log-\ac{LRT} 
\begin{equation} \label{eq:Lr}
    \ell(\boldr)= \log \frac{\pX{\boldr|\boldp}}{\pX{\boldr|\boldp+\bolddelta}} \overset{\Hone}{\underset{\Htwo}{\gtrless}} 0\, .
\end{equation}
where $\pX{\boldr|\boldp}$ is the conditional \ac{pdf} of $\boldr$.
Notably, the channels associated with the two hypotheses correspond to those the \ac{BS} would use to communicate with two different \acp{UE} located at positions $\boldp_1=[d,\theta]^{\top}$ and $\boldp_2=[d+\Delta_{\mathrm{d}},\theta+\Delta_{\theta}]^{\top}$, respectively, performing beamfocusing on the respective positions (conjugate beamforming) \cite{ZhaShlGuiDarImaEld:J22}. The minimum error
probability $\Pmin{\boldp,\boldp + \bolddelta}$ coincides with that of a binary detector discriminating between the two waveforms $\bolds(\boldp)$ and $\bolds(\boldp+\bolddelta)$, each having the same energy $\alpha^2$ under hypothesis $\Hone$ and $\Htwo$, respectively. The detector effectively operates as a correlator matched to the waveform difference $\bolds(\boldp) - \bolds(\boldp+\bolddelta)$, and its performance depends on the correlation between the two waveforms, that is, the similarity between the channels corresponding to the positions $\boldp_1$ and $\boldp_2$. 
It is well known that the optimal correlator achieves an error probability given by\footnote{We note that, differently from the ZZB for time-based distance estimation (e.g., in \cite{DecDar:C14}), here the error probability depends not only on the displacement $\bolddelta$ but also on the actual parameter value $\boldp$. This is consistent with the fact that the \ac{CRB} expressions in \eqref{CRBtradRange}-\eqref{CRBtradAOA} also depend on the parameter values $d$, $\theta$, since the estimation accuracy varies with the spatial position. In contrast, for time-based estimation the CRB, thus the estimation quality, does not depend on the specific value of the delay itself.} \cite{Skl:B01} 
\begin{equation} \label{eq:PminPMB}
    \Pmin{\boldp,\boldp+\bolddelta}=\Q{\sqrt{K \, \SNR\left ( 1-\rho(\boldp,\boldp+\bolddelta) \right)}}\,
\end{equation}
where $\Q{\cdot}$ is the Gaussian $Q$-function, and  $\rho(\boldp,\boldp+\bolddelta)$ is the normalized correlation function between $\bolds(\boldp)$ and ${\bolds(\boldp+\bolddelta)}$, given by
\begin{equation} \label{eq:rhos}
    \rho(\boldp,\boldp+\bolddelta)=\frac{1}{K} \left |\left(\bolds(\boldp+\bolddelta)\right)^\dag  \bolds(\boldp) \right |.
\end{equation}
Interestingly, adopting the \ac{ZZB}, the performance characterization of an estimation problem is put in relationship with that of a corresponding binary detection problem.

Taking into account the uniform distributions of distance and \ac{AoA} as in \eqref{eq:pdf_p}, the \ac{ZZB} in \eqref{eq:ZZB2Dgen} can be specified for distance estimation with $\bolda=[1,0]^{\top}$ as
\begin{align}\label{eq:ZZB2Ddistance}
    &\ZZB^{( \hat{d})}=   \\
    &\frac{1}{T_{\mathrm{d}} T_{\theta}}\!\int_0^{T_{\mathrm{d}}} \!\!\!\max _ { \bolddelta : \Delta_{\mathrm{d}} = h } \!\int_{\dmin}^{\dmax-h} \!\!\!\int_{\tmin}^{\tmax-\Delta_\theta} \!\!\!
 P_{\min }(\boldp, \boldp\!+\!\bolddelta) \,\dd \theta \,\dd d \, h \, \dd h\, . \nonumber
\end{align}
The \ac{ZZB} is then given by \eqref{eq:ZZB2Ddistance} with the $\Pmin{\boldp,\boldp+\bolddelta}$ expression \eqref{eq:PminPMB}.
For \ac{AoA} estimation ($\bolda=[0,1]^{\top}$) we get
\begin{align}\label{eq:ZZB2Dangle}
    &\ZZB^{( \hat{\theta})}= \\
    &\frac{1}{T_{\mathrm{d}} T_{\theta}}\! \int_0^{T_{\theta}} \!\!\!\max _ { \bolddelta : \Delta_\theta = h } \!\int_{\dmin}^{\dmax-\Delta_{\mathrm{d}}} \!\!\!\int_{\tmin}^{\tmax-h} \!\!\!
 P_{\min }(\boldp, \boldp\!+\!\bolddelta) \,\dd \theta \,\dd d \, h \, \dd h  . \nonumber
\end{align}
It is worth noticing that in both cases, the output is the \ac{ZZB} of one parameter assuming a certain prior information about both parameters, i.e., the \ac{ZZB} of one parameter in case the parameters are jointly estimated.
As an example, when considering distance estimation according to \eqref{eq:ZZB2Ddistance}, fixed an amount of error $h$ on the distance (outer integral), all the error possibilities between position $\boldp=[d, \theta]^{\top}$ and ${\boldp+\bolddelta=[d+h, \theta+\Delta_\theta]^{\top}}$ must be considered (i.e., all the possible $\Delta_\theta$), then accounting for the worst case. 

When examining the problem of distance estimation only (i.e., with \ac{AoA} $\theta$ known), it can be shown that the \ac{ZZB} becomes
\begin{equation} \label{eq:ZZBfinal1}
\ZZB^{( \hat{d};\theta)} = \frac{1}{T_{\mathrm{d}}} \int_0^{T_{\mathrm{d}}} \left\{ \int_{\dmin}^{\dmax-h} \Pmin{d,d+h} \dd d \right\} h\,  \dd h \, .
\end{equation}
where we denoted by $\Pmin{d,d+h}$ the error probability of the \ac{LRT} \eqref{eq:Lr} for the specific \ac{AoA} considered.

At low \ac{SNR}, we have $\Pmin{d,d+h}\approx \frac{1}{2}, \forall h$. Under this condition, the \ac{ZZB} in \eqref{eq:ZZBfinal1} has a value $\Td^2/12$, which is reasonable as, in the absence of any information from the observations (due to the low \ac{SNR} regime), any estimator would guess the distance starting from the a-priori knowledge only. Having considered a uniform distribution, the \ac{MSE} must then equal the variance of the uniform prior distribution.

At high \ac{SNR}, the behavior of the \ac{ZZB} is strongly dependent on the autocorrelation function $ \rho(\boldp,\boldp+\bolddelta)$ in \eqref{eq:rhos}, which is
\begin{align}
\rho(\boldp,\boldp+\bolddelta)
& = \frac{1}{K}\left| \sum_{k=-\frac{K-1}{2}}^{\frac{K-1}{2}} e^{\jm \frac{2\pi}{\lambda} \left[ d_k(d+\Delta_{\mathrm{d}},\theta+\Delta_{\theta})-d_k(d,\theta) \right] } \right| \, .
\end{align}
Assuming a transmitter on the broadside direction of the array (i.e., $\theta=0$), leading to the best accuracy for distance estimation according to \eqref{CRBtradRange}, it is possible to show that at high \ac{SNR} the \ac{ZZB} for distance estimation converges to\footnote{The demonstration is provided in the appendix.}
\begin{equation}\label{eq:asympt}
\ZZB^{( \hat{d};\theta=0)} \rightarrow \frac{ 18 \lambda^2 \left(d_{\max}^5 - d_{\min}^5\right)}{\pi^2 K\SNR \, \Da^4 \, T_{\mathrm{d}}}\, .
\end{equation}

Incidentally, \eqref{eq:asympt} corresponds to the global \ac{CRB} for distance estimation, denoted by $\CRBg^{(\hat{d})} $, i.e., what is obtained by averaging the local \ac{CRB} in \eqref{CRBtradRange} over the prior distribution of the parameters. Specifically, we have
\begin{equation}
\CRBg^{(\hat{d})} = \int_{\dmin}^{\dmax}\int_{\tmin}^{\tmax} p_{\boldp}(\boldp) \, \CRB^{(\hat{d})} \, \dd \theta\, \dd d\, 
\end{equation}
which, for $\theta=0$, returns \eqref{eq:asympt} as $\delta^4 \left ( K^2\!-\!4\right ) \left ( K^2 \!-\!1\right ) \approx \Da^4$.
When $\dmin=0$ we have, for $\theta=0$ 
\begin{equation}\label{eq:CRBglobaldistance}
\CRBg^{(\hat{d};\theta=0)} \!=\! \frac{18 c^2 \dmax^4 }{\pi^2 \!\fc^2 K \, \SNR  \, \Da^4}
\end{equation}
showing that the global \ac{CRB} has the same expression as the local \ac{CRB} evaluated for a distance $d=\dmax/\sqrt[4]{5}$ (effective distance).

\section{Numerical Results}

We consider a transmitter in position $\boldp=[d,0]^{\top}$ transmitting a narrowband signal at frequency $\fc=\unit[28]{GHz}$ that is received by an \ac{ELAA} with variable number $K$ of antennas and fixed aperture $\Da=\unit[1]{m}$. Being $\lambda\approx\unit[1]{cm}$, a standard half-wavelength spaced array requires approximately $200$ antenna elements.

\begin{comment}
\begin{figure}[t]
	\centerline{\includegraphics[width=\columnwidth]{Figures/ZZB_NF_ranging-eps-converted-to.pdf}}
	\caption{Esempio di confronto\ac{ZZB} - \ac{CRB} globale -\ac{MSE} per due diversi numeri di antenne (8 e 64) e apertura fissata (spacing variabile). Si vede come lo\ac{ZZB} preveda bene l'SNR di soglia del\ac{MSE} relativo allo stimatore ML. Ovviamente funziona meglio il caso con piu' antenne (maggiore\ac{SNR} gain ad apertura fissata nella regione asintotica in accordo a (6)). Tuttavia non e' l'unico effetto quando guardiamo lo\ac{ZZB}, come mostrato nel risultato seguente.}
	\label{fig:Example}
\end{figure}

\begin{figure}[t]
    	\centerline{\includegraphics[width=\columnwidth]{Figures/ZZB_varK-eps-converted-to.pdf}}
	\caption{ZZB normalizzato (moltiplicato per $K$) in modo da essere indipendente dall'SNR gain dovuto all'utilizzo di un diverso numero di antenne. Apertura fissata. Una volta così normalizzato, in accordo a (6), il \ac{CRB} dipende solo dall'apertura. Difatti si vede che lo\ac{ZZB} asintotico cambia poco per i diversi numeri di antenne, come atteso (a meno della validita' dell'approx). Tuttavia, l'SNR di soglia cambia. Questo perchè diminuendo il numero di antenne ad apertura fissata si allarga lo spacing e quindi crescono le ambiguita' (sidelobes della funzione di verosimiglianza). Questo fatto non e' tenuto in conto dal \ac{CRB}. Trovare una espressione per l'SNR di soglia potrebbe essere interessante per questa configurazione.}
	\label{fig:Example}
\end{figure}
\end{comment}

\begin{figure}[t]
    	\centerline{\includegraphics[width=0.99\columnwidth]{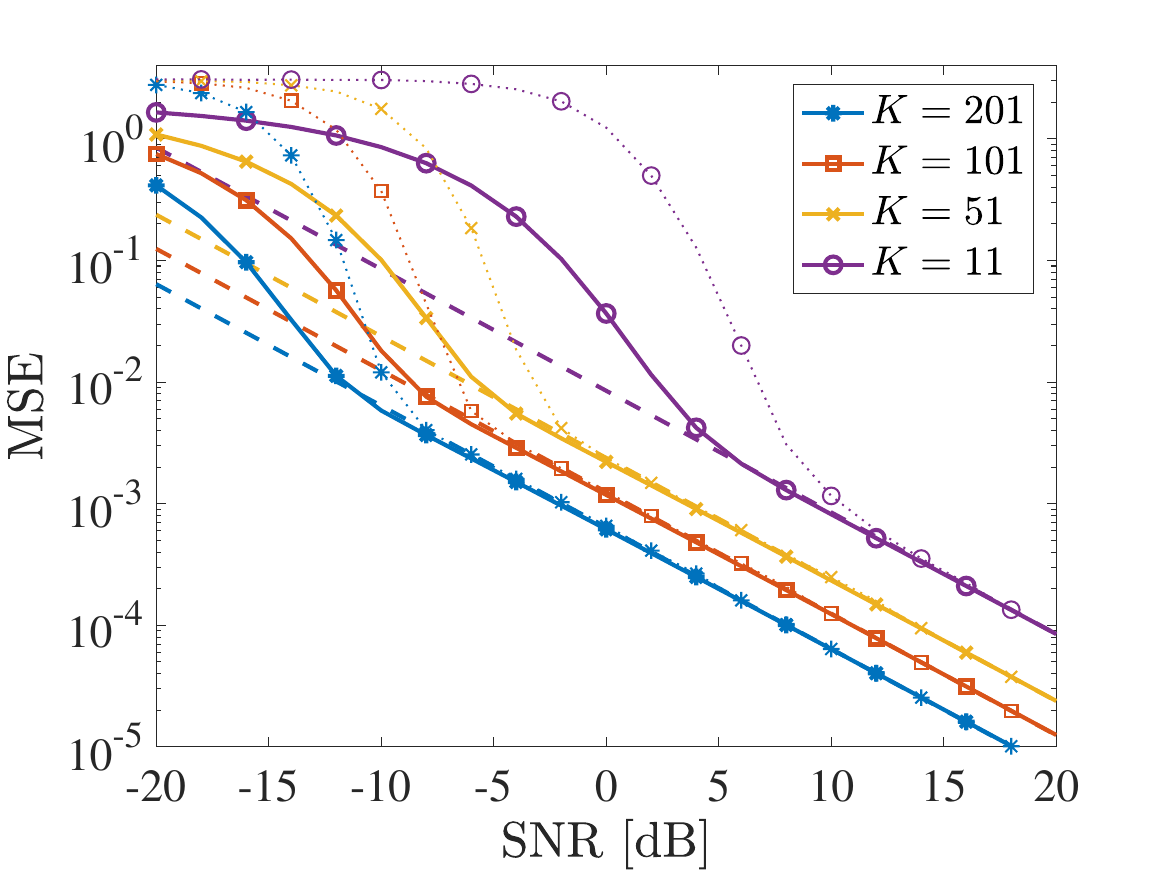}}
	\caption{ZZB for distance estimation (bold lines) as a function of the SNR for different number of antennas, with known AoA . Comparison with the CRB (-$\,$-) and with MSE of the MLE ($\cdot\cdot$).}
	\label{fig:ZZBvsSNR_NotNormalized}
\end{figure}

Figure~\ref{fig:ZZBvsSNR_NotNormalized} reports the \ac{ZZB} for distance estimation obtained with \eqref{eq:ZZBfinal1} as a function of the single-antenna \ac{SNR} for different numbers of antennas $K$, thus for different antenna spacing. The transmitter is supposed to be randomly located  in $[\dmin,\dmax]=[0,5]\,\unit{m}$. For comparison, the global \ac{CRB} in \eqref{eq:CRBglobaldistance} is reported (dashed lines). Moreover, the \ac{MSE} of the \ac{MLE} obtained using Montecarlo simulations and considering a grid-search approach is reported for comparison. It is possible to see that the accuracy degrades as the number of antennas decreases. This holds for both the \ac{ZZB} and the \ac{CRB}. In fact, according to \eqref{eq:CRBglobaldistance}, even for fixed aperture $\Da$, as in this case, increasing the number of antennas leads to a lower \ac{CRB} thanks to the increased number of independent observations (i.e., \ac{SNR} gain). For low \ac{SNR}, all the \ac{ZZB} curves converge to the value $\unit[2]{m^2}$, which corresponds to $\dmax^2/12$, as expected. Differently, at high \ac{SNR}, the \ac{ZZB} converges to the corresponding global \ac{CRB}. At intermediate \ac{SNR}, when ambiguities cause larger estimation errors, the \ac{ZZB} starts to diverge from the \ac{CRB}. The \ac{MLE} approaches the \ac{CRB} at high \ac{SNR}, confirming its asymptotic efficiency. The \ac{ZZB} is tighter with the actual estimator performance for a wide range of \ac{SNR} and, in particular, it can well predict the so-called \textit{\ac{SNR} threshold} where the performance of the estimation starts deviating from the \ac{CRB}. The \ac{SNR} threshold changes for the different configurations. In particular, the \ac{SNR} threshold increases as the number of antennas decreases. This can be explained because reducing the number of antennas increases the level of sidelobes and widens the ambiguity zones in the waveform correlation. Thus, the asymptotic performance is more difficult to achieve as the antenna spacing grows over half-wavelength.

\begin{figure}[t]
    	\centerline{\includegraphics[width=0.99\columnwidth]{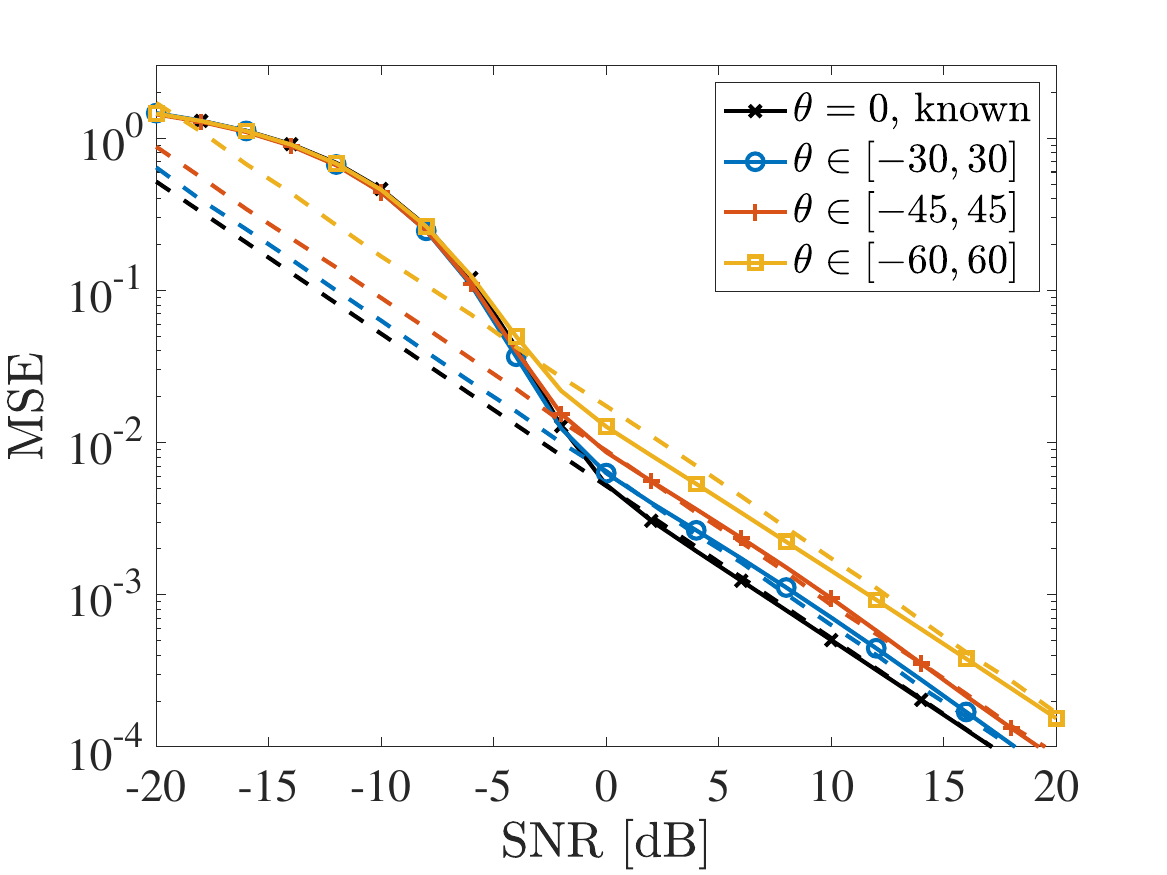}}
	\caption{ZZB for distance estimation as a function of the SNR for $K=21$ with unknown AoA span $[\tmin, \tmax]$. Comparison with the CRB (-\,-).}
	\label{fig:ZZBvsSNRdistanceangle}
\end{figure}

Figure~\ref{fig:ZZBvsSNRdistanceangle} reports the \ac{ZZB} for distance estimation as a function of the \ac{SNR} when distance and \ac{AoA} are considered jointly. 
In this case, the \ac{ZZB} for distance estimation is obtained using \eqref{eq:ZZB2Ddistance} for different \ac{AoA} spans $[\tmin, \tmax]$ and $K=21$. The \ac{ZZB} for known angle $\theta=0$ obtained with \eqref{eq:ZZBfinal1} is also reported for comparison. As before, at low \ac{SNR} the \ac{ZZB} converges to $\dmax^2/12$. At high \ac{SNR} the \ac{ZZB} is very close to the corresponding global \ac{CRB} obtained by averaging the local \ac{CRB} expression \eqref{CRBtradRange} over the \ac{AoA} and distance prior distribution in \eqref{eq:pdf_p} (see dashed lines). Here, as the angular span increases, the corresponding performance decreases since estimation accuracy gets worse for large angles (i.e., transmitter far from the antenna broadside direction, where phase variations coming from the spherical wavefront propagation are more evident).
Interestingly, despite the fact that now the \ac{AoA} $\theta$ is supposed unknown, it is possible to observe that the \ac{SNR} threshold  is practically unaffected by different \ac{AoA} spans. This can be ascribed to the fact that, with sufficiently large arrays, the estimation of the \ac{AoA} is extremely accurate even at moderate \ac{SNR}, 
so that it does not worsen global ambiguities.

\section{Conclusion}

In this work, we analyzed the \ac{ZZB} for near-field localization with large antenna arrays, offering a tighter characterization of estimation performance compared to the \ac{CRB}. 
We derived the asymptotic behavior of the \ac{ZZB} in low- and high-\ac{SNR} regimes, and analytically proved that the bound converges to the global \ac{CRB} at high \ac{SNR}. We also showed that, for fixed array aperture, decreasing the number of antennas alters the \ac{SNR} threshold due to ambiguity effects when considering near-field distance estimation. Moreover, we found that the \ac{SNR} threshold for distance estimation is only weakly influenced by the quality of the prior angle information.

Overall, these results clarify the fundamental limits and threshold behavior of near-field localization and sensing and can guide the design and performance evaluation of next-generation array processing techniques.

\bigskip
\appendix
When $\theta=0$, by adopting the Fresnel approximation\footnote{The approximation is known to be tight for $d>\Da$ \cite{GioDecZanDar:C25}.} for the term $d_k(d,\theta)$, i.e., $d_k(d,\theta)\approx d + \frac{x_k^2}{2d}$, we obtain
\begin{align}
\rho(d,d+h)&\approx\frac{1}{K}\left| \sum_{k=-\frac{K-1}{2}}^{\frac{K-1}{2}} e^{\jm \frac{2\pi}{\lambda} \left[ \bar{d} + \frac{x_k^2}{2\bar{d}} -d -\frac{x_k^2}{2d} \right]}
\right|  
\end{align}
where we used the notation $\rho(d,d+h)$ to indicate $\rho(\boldp,\boldp+\bolddelta)$ for $\theta=0$ and $\Delta_{\theta}=0$, and with $\bar{d}=d+h$.
By neglecting the common phase terms that do not contribute to the behavior of the cross-correlation function, and using $x_k=k \delta$ we have
\begin{align} \label{eq:rhoA}
\rho(d,d+h)&\approx\frac{1}{K}\left| \sum_{k=-\frac{K-1}{2}}^{\frac{K-1}{2}} e^{-\jm \frac{\pi}{\lambda} k^2\delta^2 \frac{h}{d(d+h)}}
\right| .
\end{align}
We now approximate the sum in \eqref{eq:rhoA} with the corresponding Riemann integral, obtaining
\begin{align} \label{eq:rhoB}
\rho(d,d\!+\!h)&\!\approx\!\frac{1}{K}\left| \int_{-\frac{K-1}{2}}^{\frac{K-1}{2}}\!\!\!e^{-\jm \frac{\pi}{\lambda} k^2\delta^2 z }  \dd k \right| \!=\!\frac{2}{K}\left| \int_{0}^{\frac{K-1}{2}} \!\!\!e^{-\jm \frac{\pi}{\lambda} k^2\delta^2 z }  \dd k \right| 
\end{align}
where $z=\frac{h}{d(d+h)}$ and the last equality holds since the integrand is an even function of $k$. Using a change of variable $\xi=\sqrt{\frac{2k^2\delta^2 z}{\lambda}}$ it is obtained
\begin{align} \label{eq:rhoC}
\rho(d,d+h)&\approx\frac{2}{K}\left| \int_0^{\nu} \sqrt{\frac{\lambda}{2\delta^2 z}}e^{-\jm\frac{\pi}{2}\xi^2} \dd \xi \right| 
\end{align}
where $\nu=\frac{K-1}{2}\sqrt{\frac{2\delta^2 z}{\lambda}}\approx\sqrt{\frac{K^2\delta^2 z}{2\lambda}}$ (large number of antennas, thus $K-1 \approx K$). Equivalently, \eqref{eq:rhoC} can be written as
\begin{align} \label{eq:rhoD}
&\rho(d,d+h)\approx\nonumber \\
&\,\,\sqrt{\frac{2\lambda}{K^2\delta^2 z}}\left| \int_0^{\nu}\cos{\left(\frac{\pi}{2}\xi^2\right)} \dd \xi -\jm \int_0^{\nu}\sin{\left(\frac{\pi}{2}\xi^2\right)} \dd \xi \right| \nonumber \\
&\quad=\sqrt{\frac{C^2(\nu) + S^2(\nu)}{\nu^2}}
\end{align}
where $C(u)=\int_0^u \cos{\left(\frac{\pi}{2}t^2\right)} \dd t$ and $S(u)=\int_0^u\sin{\left(\frac{\pi}{2}t^2\right)} \dd t$ are the Fresnel integrals.

In order to find the behavior of the \ac{ZZB} for large \ac{SNR}, the values of $\rho(d,d+h)$ of interest are that corresponding to small values of $h$, since the possibility of estimation is guaranteed by the ability to distinguish between a transmitter located at distance $d$ from that located immediately close (behavior of the cross-correlation function around the main lobe peak). To this end, we can compute the Taylor expansion of $\rho(d,d+h)$ for small $h$, obtaining
\begin{align} \label{eq:rhoE}
\rho(d,d+h)&\approx1-\frac{\pi^2}{90}\nu^4 \approx1-\frac{\pi^2\delta^4K^4h^2}{360\lambda^2d^4}
\end{align}
where in the right-hand approximation we used $d(d+h)\approx d^2$.
Thus, the minimum error probability in \eqref{eq:PminPMB} is given, for small $h$, by 
\begin{align} \label{eq:PminA}
    \Pmin{d,d+h}  \approx \Q{\sqrt{ \frac{K\SNR\, \pi^2 \Da^4}{360 \lambda^2 d^4}}h}
\end{align}
where we used $\delta K \approx \Da$. 

When $\SNR \gg 1$ (large \ac{SNR}), we evaluate the \ac{ZZB} integral in \eqref{eq:ZZBfinal1}, denoted by $I$ in the following, using \eqref{eq:PminA}, which is
\begin{equation} 
I = \frac{1}{T_{\mathrm{d}}} \int_0^{T_{\mathrm{d}}} \left\{ \int_{\dmin}^{\dmax-h} \Q{\frac{\gamma h}{d^2}} \dd d \right\} h\,  \dd h \, 
\end{equation}
with $\gamma=\sqrt{ \frac{K\SNR \pi^2 \Da^4}{360 \lambda^2}}$.
In fact, for large $\gamma$, the main contribution to the integral comes from small $h$, corresponding to the region of validity of the previous Taylor expansion for the cross-correlation function. Here, the integrand $Q(\gamma h / d^2)$ is only significant for $h \ll T_{\mathrm{d}}$, so that the upper limit $d_{\max}-h$ can be approximated as $d_{\max}$.
Since the integral is dominated by $h \ll T_{\mathrm{d}}$, by exchanging the order of integration and extending the upper limit of $h$ to infinity, we  obtain
\begin{equation}
I \approx \frac{1}{T_{\mathrm{d}}} \int_{d_{\min}}^{d_{\max}} 
 \int_0^{\infty} h \, Q\left( \frac{\gamma h}{d^2} \right) \dd h \,  \dd d \, .
\end{equation}
For a fixed $d$, we introduce the change of variable $u = \frac{\gamma h}{d^2}$.  Then the inner integral becomes
\begin{align}
\int_0^{\infty} h \, Q\left( \frac{\gamma h}{d^2} \right) \dd h 
&= \int_0^{\infty} \frac{d^2}{\gamma} u \, Q(u)  \frac{d^2}{\gamma} \dd u \nonumber \\
&= \frac{d^4}{\gamma^2} \int_0^{\infty} u \, Q(u) \, \dd u = \frac{d^4}{4 \gamma^2}.
\end{align}
Finally, we get
\begin{align}
I &\approx \frac{1}{T_{\mathrm{d}}} \int_{d_{\min}}^{d_{\max}} \frac{d^4}{4 \gamma^2}  \, \dd d 
= \frac{d_{\max}^5 - d_{\min}^5}{20 \, \gamma^2 \, (d_{\max}-d_{\min})
} 
\end{align}
corresponding to the global \ac{CRB} for distance estimation in \eqref{eq:asympt}.

\bigskip 
\section*{Acknowledgment}
This work was partially supported by the European Union under the Italian National Recovery and Resilience Plan (NRRP) of 
NextGenerationEU, partnership on “Telecommunications of the Future” (PE00000001 - program “RESTART”).

\bigskip 
\bibliographystyle{IEEEtran}
\bibliography{Biblio/IEEEabrv,Biblio/Stringdefinitions,Biblio/BiblioDD,Biblio/biblio,Biblio/WINS-Books}

\end{document}

%% file: acronyms.tex
\begin{acronym} 
\acro{3GPP}{Third Generation Partnership Project}
\acro{2D}{two-dimensional}
\acro{2G}{second generation}
\acro{3G}{third generation}
\acro{4G}{fourth generation}
\acro{5G}{fifth generation}
\acro{6G}{sixth generation}
\acro{5GAA}{5G Automotive Association}
\acro{5GS}{5G system}
\acro{AF}{application function}
\acro{AMC}{adaptive modulation and coding}
\acro{AS}{application server}
\acro{AoA}{angle of arrival}
\acro{AWGN}{additive white Gaussian noise}
\acro{AV}{automated vehicle}
\acro{BEP}{beacon error probability}
\acro{BM-SC}{Broadcast Multicast Service Center}
\acro{BP}{beacon periodicity}
\acro{B-CSA}{broadcast \ac{CSA}}
%\acro{BR}{beacon resource}
%\acro{BRF}{\ac{BR}-frequency}
%\acro{BRT}{\ac{BR}-time}
\acro{BS}{base station}
\acro{BSM}{basic safety message}
\acro{BSS}{basic service set}
\acro{C-ITS}{cooperative-intelligent transport systems}
\acro{C-NOMA}{Cooperative \ac{NOMA}}
\acro{C-V2X}{cellular-\ac{V2X}}
\acro{CA}{collision avoidance}
\acro{CAM}{cooperative awareness message}
\acro{CAV}{connected automated vehicle}
\acro{CBR}{channel busy ratio}
\acro{cdf}{cumulative distribution function}
\acro{ccdf}{complementary cumulative distribution function}
\acro{CDMA}{code-division multiple access}
\acro{CFAR}{constant false alarm rate}
\acro{CoMP}{coordinated multi-point}
\acro{CN}{core network}
\acro{CP}{cyclic prefix}
\acro{CP-OFDM}{cyclic prefix \ac{OFDM}}
\acro{CPM}{collective perception message}
\acro{CR}{channel occupancy ratio}
\acro{CRB}{Cramér-Rao bound }
\acro{CRDSA}{contention resolution diversity slotted ALOHA}
\acro{CSA}{coded-slotted ALOHA}
\acro{CSD}{cyclic shift diversity}
\acro{CSI}{channel state information}
\acro{CSIT}{channel state information at the transmitter}
\acro{CSIR}{channel state information at the receiver}
\acro{CSMA/CA}{carrier sense multiple access with collision avoidance}
\acro{D2D}{device-to-device}
\acro{DCC}{decentralized congestion control}
\acro{DCF}{distributed coordination function}
\acro{DCI}{downlink control
information}
\acro{DCM}{Dual carrier modulation}
\acro{DENM}{decentralized environmental notification message}
\acro{DMRS}{demodulation reference signal}
\acro{DRX}{discontinuous reception}
\acro{DSRC}{dedicated short range communication}
\acro{DOT}{Department of Transportation}
\acro{EC}{European Commission}
\acro{ECP}{extended cyclic prefix}
\acro{EDCA}{enhanced distributed coordination access}
\acro{EDGE}{Enhanced Data rates for GSM Evolution}
\acro{ELAA}{extremely large aperture array}
\acro{eMBMS}{evolved Multicast Broadcast Multimedia Service}
\acro{eMMB}{enhanced mobile broadband}
\acro{eNodeB}{evolved NodeB}
\acro{EPC}{Evolved Packet Core}
\acro{EPS}{Evolved Packet System}
\acro{ETSI}{European Telecommunications Standards Institute}
\acro{FCC}{Federal Communications Commission}
\acro{FCD}{floating car data}
\acro{FD}{in-band full-duplex}
\acro{FDD}{frequency division duplex}
\acro{FDMA}{frequency division multiple access}
\acro{FEC}{forward error correction}
\acro{FIM}{Fisher information matrix}
\acro{FFT}{fast Fourier transform}
\acro{GDOP}{geometric dilution of precision}
\acro{GNSS}{global navigation satellite system}
\acro{GPRS}{general packet radio service }
\acro{GPS}{global positioning system}
\acro{HARQ}{hybrid automatic repeat request}
\acro{HD}{half-duplex}
\acro{HSDPA}{High Speed Downlink Packet Access}
\acro{HSPA}{High Speed Packet Access}
\acro{IBE}{in-band emission}
\acro{ICI}{Inter-carrier interference}
\acro{IMU}{inertial measurement unit}
\acro{IDMA}{interleave-division multiple access}
\acro{IFFT}{inverse fast Fourier transform}
\acro{i.i.d.}{independent identically distributed}
\acro{IM}{index modulation}
\acro{IP}{Internet Protocol}
\acro{IPG}{inter-packet gap}
\acro{IR}{incremental redundancy}
\acro{ISAC}{integrated sensing and communication}
\acro{ISI}{inter-symbol interference}
\acro{ITS}{intelligent transport system}
\acro{i.i.d.}{independent identically distributed}
\acro{JCS}{joint communication and sensing}
\acro{KPI}{key performance indicator}  
\acro{LDPC}{low-density parity-check}
\acro{LDS}{low-density spreading}
\acro{LiDAR}{light detection and ranging}
\acro{LOS}{line-of-sight}
\acro{LRT}{likelihood ratio test}  
\acro{LTE}{long term evolution}  
\acro{LTE-D2D}{\ac{LTE} with \ac{D2D} communications}  
\acro{LTE-V2X}{long-term-evolution-vehicle-to-anything} \acro{LTE-V2V}{\ac{LTE}-\ac{V2V}}
\acro{LAR}{localization availability ratio}  
\acro{LUR}{localization update rate}  
\acro{MAC}{medium access control}
\acro{MBMS}{Multicast Broadcast Multimedia Service}
\acro{MBMS-GW}{MBMS Gateway}
\acro{MBSFN}{MBMS Single Frequency Network}
\acro{mmWave}{millimeter-wave}
\acro{MCE}{Multi-cell Coordination Entity}
\acro{MEC}{mobile edge computing}
\acro{MCM}{maneuver coordination message}
\acro{MCS}{modulation and coding scheme}
\acro{MIMO}{multiple input multiple output}
\acro{ML}{maximum likelihood}
\acro{MLE}{maximum likelihood estimator}
\acro{MRD}{maximum reuse distance}
\acro{MSE}{mean squared error}
\acro{MUD}{multi-user detection}
\acro{NEF}{network exposure function}
\acro{NGV}{Next Generation V2X}
\acro{NHTSA}{National Highway Traffic Safety Administration}
\acro{NLOS}{non-line-of-sight}
\acro{NOMA}{non-orthogonal multiple access}
\acro{NOMA-MCD}{\ac{NOMA}-mixed centralized/distributed}
\acro{NV}{neighbor vehicle}
\acro{NR}{new radio}
\acro{PEB}{position error bound}
\acro{OBU}{on-board unit}
%\acro{OCB}{outside of the context of a \ac{BSS}}
\acro{OCB}{outside of the context of a basic service set}
\acro{OEM}{Original equipment manufacturer}
\acro{OFDM}{orthogonal frequency-division multiplexing}
\acro{OFDMA}{orthogonal frequency-division multiple access}
\acro{OMA}{orthogonal multiple access}
\acro{OTFS}{orthogonal time frequency space}
\acro{pdf}{probability density function}
\acro{PAPR}{peak to average power ratio}
\acro{PCF}{policy control function}
\acro{PCM}{platoon control message}
%\acro{PD}{packet delay}
\acro{PoD}{probability of detection}
\acro{PoFA}{probability of false alarm}
\acro{PEP}{pairwise error probability} 
\acro{PER}{packet error rate}
\acro{PIAT}{packet inter-arrival time}
\acro{PSCCH}{Physical sidelink control channel}
\acro{PSFCH}{Physical sidelink feedback channel}
\acro{PSSCH}{Physical sidelink shared channel}
\acro{PSBCH}{Physical sidelink broadcast channel}
\acro{PHY}{physical}
\acro{PL}{path loss}
\acro{PLMN}{Public Land Mobile Network}
\acro{PPPP}{ProSe Per-Packet Priority}
\acro{PPPR}{ProSe Per-Packet Reliability}
\acro{ProSe}{Proximity-based Services}
\acro{PRB}{physical resource block}
\acro{PRR}{packet reception ratio}
\acro{PSD}{power spectral density}
\acro{QAM}{quadrature amplitude modulation}
%\acro{QC-LDPC}{quasi-cyclic low-density parity-check}
\acro{QC-LDPC}{quasi-cyclic \ac{LDPC}}
\acro{QoS}{quality of service}
\acro{QPSK}{quadrature phase shift keying}
\acro{RADAR}{radio detection and ranging}
\acro{RAN}{radio access network}
\acro{RAT}{radio access technology}
\acro{RB}{resource block}
\acro{RBP}{resource block pair}
\acro{RCS}{radar cross section}
\acro{RE}{resource element}
\acro{REB}{ranging error bound}
\acro{RF}{radio frequency}
\acro{RIS}{reconfigurable intelligent  surface}
\acro{ROC}{receiver operating characteristic}
\acro{RR}{radio resource}
\acro{RRC}{radio resource control}
\acro{RSRP}{reference signal received power}
\acro{RSSI}{received signal strength indicator}
\acro{RSU}{road side unit} 
\acro{RTT}{round trip time} 
\acro{R.V.}{random variable}
\acro{SAI}{Service Area Identifier}
\acro{SC-FDMA}{single carrier frequency division multiple access}
\acro{SC-PTM}{Single Cell Point To Multipoint}
\acro{SCS}{subcarrier spacing}
\acro{SCMA}{sparse-code multiple access}
\acro{SI}{self-interference}
\acro{SIC}{successive interference cancellation}
\acro{S-UE}{scheduling UE}
\acro{SCI}{sidelink control information}
\acro{SDN}{Software-defined networking}
\acro{SFFT}{symplectic finite Fourier transform}
\acro{SHINE}{simulation platform for heterogeneous interworking networks}
\acro{SIMO}{single input multiple output}
\acro{SINR}{signal-to-interference-plus-noise ratio}
\acro{SISO}{single input single output}
\acro{SNR}{signal-to-noise ratio}
\acro{SPS}{semi-persistent scheduling}
\acro{SRS}{sounding reference signal}
\acro{STBC}{space time block codes}
\acro{SV}{smart vehicle}
\acro{TB}{transport block}
\acro{TBC}{time before change}
\acro{TBE}{time before evaluation}
\acro{TDD}{time-division duplex}
\acro{TDMA}{time-division multiple access}
\acro{TDoA}{time-difference-of-arrival}
\acro{ToA}{time-of-arrival}
\acro{ToF}{time-of-flight}
\acro{TNR}{treshold to noise ratio}
\acro{TTI}{transmission time interval}
\acro{TWR}{two-way ranging}
\acro{UAV}{unmanned aerial vehicle}
\acro{ULA}{uniform linear array}
\acro{UD}{Update delay}
\acro{UE}{user equipment}
\acro{UMTS}{universal mobile telecommunications system}
\acro{URLLC}{ultra-reliable and ultra-low latency communications}
\acro{USIM}{Universal Subscriber Identity Module}
\acro{UTDOA}{uplink time difference of arrival}
\acro{UWB}{ultra-wide bandwidth}
\acro{V2C}{vehicle-to-cellular} 
\acro{V2N}{vehicle-to-network}
\acro{V2I}{vehicle-to-infrastructure} 
\acro{V2P}{vehicle-to-pedestrian}
\acro{V2R}{vehicle-to-roadside}
\acro{V2V}{vehicle-to-vehicle} 
\acro{V2X}{vehicle-to-everything} 
\acro{VLC}{visible light communication} 
\acro{VRU}{vulnerable road user}
\acro{VUE}{vehicular user equipment}
\acro{WAVE}{wireless access in vehicular environment}
\acro{WBSP}{wireless blind spot probability}
\acro{ZZB}{Ziv-Zakai bound}
\end{acronym}

%% file: macros.tex
\newcommand{\rednote}[1] {{\color{red}$\blacktriangleright${#1}$\blacktriangleleft$}}
\newcommand{\TBD}{\rednote{TBD}}

\newcommand{\bluenote}[1] {{\color{blue}$\blacktriangleright${#1}$\blacktriangleleft$}}

\def\erfc{{\text{erfc}}}
\def\erf{{\text{erf}}}

%Special Symbols************

\newcommand{\rank}{{\rm rank}}
\newcommand{\diag}{{\rm diag}}

\newcommand{\Matrix}{\underline{\underline{M}}}
\newcommand{\Sim}{\underline{\underline{Q}}}
\newcommand{\Graph}{G(\mathbf{N},\mathbf{A})}
\newcommand{\Fim}{\underline{\underline{J}}}
\newcommand{\Vr}{\text{Var}(\varepsilon)}
\newcommand{\Simt}{\underline{\underline{Q}}^{T}}
%\newcommand{\arg}{{\rm arg}}

% \DeclareMathOperator{\arccot}{arccot} 

%%%%%% GENERAL %%%%%%

\newtheorem{theorem}{Theorem}
\newtheorem{definition}{Definition}
\newtheorem{corollary}{Corollary}
\newtheorem{lemma}{Lemma}

\newcommand{\TNR}{\mathsf{TNR}}
\newcommand{\SNR}{\mathsf{SNR}}
\newcommand{\PEB}{\mathsf{PEB}}

\newcommand{\tr}{^{\text{T}}}

\def\erfc{{\text{erfc}}}
\def\erf{{\text{erf}}}
\def\inve{{\text{inverfc}}}
\def\teq{\triangleq}
\def\bs{$\blacksquare$}
\newcommand{\Q}[1] {Q \left ( #1 \right )}

\newcommand{\rect}[1] {\text{rect} \left ({#1} \right )}
\newcommand{\sinc}[1] {\text{sinc} \left ({#1} \right )}

\newcommand{\Real}[1]{\,\Re\left\{#1\right\}}
\newcommand{\floor}[1] {f \left ({#1} \right )}

\newcommand{\err} {e\left( \pos \right)}
\newcommand{\errx} {e_{\mathrm{th}}}%{e^\star}
\newcommand{\errm} {e_{m}\left( \pos \right)}
\newcommand{\errmsq} {e_{m}^{2}\left( \pos \right)}
\newcommand{\errLB} {e_{\text{LB}} \left( \pos \right)}
\newcommand{\errLBq} {e_{\text{LB}}^2 \left( \pos \right)}

\newcommand{\Hk}{{\mathcal{H}}}
\newcommand{\Hak}{{\mathcal{D}}}
\newcommand{\Bk}{{\mathcal{B}}}
\newcommand{\Rate}{R}

\newcommand{\PX}[1] {{\mathbb{P}}\left\{{#1}\right\}}
\newcommand{\EX}[1] {{\mathbb{E}}\left\{{#1}\right\}}
\newcommand{\Var}[1] {{\mathbb{V}}\left\{{#1}\right\}}\newcommand{\EXs}[2] {{\mathbb{E}}_{{#1}}\!\!\left\{{#2}\right\}}
\newcommand{\IX}[1] {{\mathbb{I}}\left\{{#1}\right\}}
\newcommand{\ONE}[2] {{\mathbbm{1}}_{#1}\left({#2}\right)}
\newcommand{\DX}[1] {{\mathbb{D}}\left\{{#1}\right\}}
\newcommand{\g}[1] {\gamma_{\mathrm{#1}}}
\newcommand{\Z}[1] {Z_{\mathrm{#1}}}
\newcommand{\V}[1] {\mathrm{var}\left\{#1\right\}}
\newcommand{\E}[1] {\mathbb{E}\left\{#1\right\}}
\newcommand{\F}[2]{\digamma^{({#2})} \left [ {#1}  \right ]}
\newcommand{\boldsped}[1] {{\bf s}_{#1}}
\newcommand{\td}[1] {\tilde{#1}}
\newcommand{\blt}[1] {\text{\boldmath $\td{\lambda}$}_{#1}}
\newcommand{\bb}[1] {\text{\boldmath ${\beta}$}_{#1}}
\newcommand{\Bv}[2] {\text{$\beta$}_{#1,#2} }
\newcommand{\vg}[1] {{\mbox{{\boldmath ${#1}$}}}}
\newcommand{\vgs}[2] {\vg{#1}_{#2}}
\newcommand{\No}{N_{0}}

\newcommand{\pd}{P_{\text{det}}}
\newcommand{\pds}{P_{\text{det}}^{\star}}
\newcommand{\Pfa}{P_{\text{fa}}}
\newcommand{\pfa}{p_{\text{fa}}}
\newcommand{\pfas}{P_{\text{fa}}^{\star}}

\newcommand{\sigmaN} {\sigma_{\text{N}}}
\newcommand{\ed} {\epsilon_{\mathrm{d}}}

\newcommand{\dd}{\text{d}}

\newcommand{\lr}{L_R}
\newcommand{\lt}{L_T}

\newcommand{\CRB}{\mathsf{CRB}}
\newcommand{\MSE}{\mathsf{MSE}}
\newcommand{\ZZB}{\mathsf{ZZB}}
\newcommand{\CRBg}{\mathsf{CRB}_{\mathrm{g}}}

\newcommand{\dmin}{d_{\mathrm{min}}}
\newcommand{\dmax}{d_{\mathrm{max}}}
\newcommand{\tmin}{\theta_{\mathrm{min}}}
\newcommand{\tmax}{\theta_{\mathrm{max}}}

\newcommand{\vd}{v_{d}}
\newcommand{\vr}{v_{\mathrm{r}}}
\newcommand{\vt}{v_{\mathrm{t}}}
\newcommand{\vk}{v_{k}}
\newcommand{\vdk}{v_{d,k}}
\newcommand{\vrk}{v_{\mathrm{r}k}}
\newcommand{\vtk}{v_{\mathrm{t}k}}
\newcommand{\pk}{p_{k}}
\newcommand{\qk}{q_{k}}

\newcommand{\rr}{{\bf r}}
\newcommand{\rt}{{\bf s}}

\newcommand{\dff}{d_{\mathrm{F}}}

\newcommand{\rhos} {\rho_{\text{s}}}
\newcommand{\rhop} {\rho_{\text{p}}}
\newcommand{\rhog} {\rho_{\text{g}}}
\newcommand{\pX}[1] {{{p}}\left\{{#1}\right\}}

\newcommand{\argmax}[1]{\underset{{#1}}{\operatorname{argmax}}}
\newcommand{\argmin}[1]{\underset{{#1}}{\operatorname{argmin}}}

\newcommand{\boldr} {{\bf r}}
\newcommand{\bolda} {{\bf a}}
\newcommand{\boldp} {{\bf p}}
\newcommand{\boldn} {{\bf n}}
\newcommand{\bolds} {{\bf s}}
\newcommand{\boldm} {{\bf m}}
\newcommand{\boldx} {{\bf x}}
\newcommand{\boldy} {{\bf y}}
\newcommand{\boldz} {{\bf z}}
\newcommand{\boldb} {{\bf b}}
\newcommand{\boldepsilon} {{\boldsymbol{\epsilon}}}
\newcommand{\bolddelta} {{\boldsymbol{\delta}}}

\newcommand{\boldgamma} {{\boldsymbol{\gamma}}}

\newcommand{\boldH} {{\bf H}}
\newcommand{\boldHa} {{\bf H}^{(\tau)}}
\newcommand{\boldHb} {{\bf H}^{(\tau+z)}}
\newcommand{\Ht} {{\bf H}_{2}}
\newcommand{\Htt} {{ \tilde{\bf{H}}}_{{2}}}
\newcommand{\boldtheta} {{\bf \theta}}
\newcommand{\hmna} {h_{m,n}^{(\tau)}}
\newcommand{\hmnb} {h_{m,n}^{(\tau+z)}}
\newcommand{\boldG} {{\bf G}}
\newcommand{\boldS} {{\bf S}^{(\tau)}}
\newcommand{\boldPH} {{\bf P}_{\boldH}}
\newcommand{\boldPHa} {{\bf P}_{\boldHa}}
\newcommand{\boldPHb} {{\bf P}_{\boldHb}}
\newcommand{\boldPHo} {{\bf P}_{\Ho}}
\newcommand{\boldPHt} {{\bf P}_{\Ht}}
\newcommand{\boldPHoo} {{\bf P}_{\Ho^{\perp}}}
\newcommand{\boldQHa} {{\boldsymbol{\mathrm{O}}}_{\boldHa}}
\newcommand{\boldQHb} {{\boldsymbol{\mathrm{O}}}_{\boldHb}}
\newcommand{\Qo} {{\bf Q}_{1}}
\newcommand{\Qt} {{\bf Q}_{2}}
\newcommand{\Qoo} {{\tilde{\bf Q}}_{1}}
\newcommand{\Qtt} {{\tilde{\bf Q}}_{2}}

%%%%%% COMMON %%%%%%

\newcommand{\Tp}{T_{\mathrm{p}}}
\newcommand{\Tc}{T_{\mathrm{c}}}
\newcommand{\Tcp}{T_{\mathrm{cp}}}
\newcommand{\Tg}{T_{\mathrm{g}}}
\newcommand{\Ts}{T_{\mathrm{sym}}}
\newcommand{\Tr}{T_{\mathrm{r}}}
\newcommand{\Th} {T_{\text{h}}}
\newcommand{\Td}{T_{\mathrm{d}}}
\newcommand{\Tf} {T_{\mathrm{f}}}
\newcommand{\Tw} {T_{\mathrm{w}}}
\newcommand{\Ta} {T_{\mathrm{a}}}
\newcommand{\Ti} {T_{\mathrm{i}}}
\newcommand{\Tb} {T_{\mathrm{b}}}
\newcommand{\Tobs}{T_{\mathrm{obs}}}
\newcommand{\Ted}{T_{\mathrm{ED}}}

\newcommand{\Da}{D_{\mathrm{a}}}

\newcommand{\Ep}{E_{\mathrm{p}}}
\newcommand{\Eb}{E_{\mathrm{b}}}
\newcommand{\Es}{E_{\mathrm{s}}}
\newcommand{\Ec} {E_{\mathrm{c}}}
\newcommand{\Et} {E_{\mathrm{t}}}
\newcommand{\Er} {E_{\mathrm{r}}}
\newcommand{\Ew} {E_{\mathrm{w}}}

\newcommand{\Ns}{N_{\mathrm{s}}}
\newcommand{\Nt}{N_{\mathrm{t}}}
\newcommand{\Nh}{N_{\mathrm{h}}}
\newcommand{\Na}{N_{\mathrm{a}}}
\newcommand{\Nr} {N_{\mathrm{r}}}
\newcommand{\Nb}{N_{\mathrm{bin}}}
\newcommand{\Nob}{N_{\mathrm{ob}}}
\newcommand{\Nu}{N_{\mathrm{u}}}
\newcommand{\Nps} {N_{\mathrm{ps}}}
\newcommand{\Nc} {N_{\mathrm{c}}}
\newcommand{\Npc} {N_{\mathrm{pc}}}
\newcommand{\Nit} {N_{\mathrm{it}}}
\newcommand{\Ne} {N_{\mathrm{w}}(i,j)}
\newcommand{\Nw} {N_{\mathrm{w}}}
\newcommand{\NA} {N_{\mathrm{b}}}
\newcommand{\NT} {N_{\mathrm{T}}}
\newcommand{\Ka}{K}
\newcommand{\Np}{N_{\mathrm{p}}}
\newcommand{\Mp}{M_{\mathrm{p}}}

\newcommand{\Nsub}{N_{\text{sub}}}
\newcommand{\Nprb}{N_{\text{ps}}}

\newcommand{\kc}{k_{\mathrm{c}}}

\newcommand{\fc} {f_{\mathrm{c}}}

\newcommand{\Ptx} {P_{\mathrm{T}}}
\newcommand{\Prx} {P_{\mathrm{R}}}

\newcommand{\jm} {\mathrm{j} }

\newcommand{\df} {\Delta\! f}

\newcommand{\yc}{y_{\mathrm{c}}}
\newcommand{\kb}{k_{\text{B}}}

\newcommand{\sn}{\sigma^2_{\mathrm{n}}}
\newcommand{\sr}{\sigma^2_{\mathrm{n}}}

\newcommand{\sone}{\star}
\newcommand{\stwo}{\star\star}
\newcommand{\sthree}{\star\star\star}

\newcommand{\Ho}{\mathcal{H}_{0}}
\newcommand{\Hone}{\mathcal{H}_{1}}
\newcommand{\Htwo}{\mathcal{H}_{2}}

\newcommand{\Pmin}[1] {{P_{\mathrm{min}}}\left ({#1}\right )}

\newcommand{\boldX}{\mathbf{X}}
\newcommand{\boldY}{\mathbf{Y}}
\newcommand{\boldR}{\mathbf{R}}
\newcommand{\boldP}{\mathbf{E}}
\newcommand{\boldI} {{\bf I}}

\newcommand{\Jrr}{J_{\vr\vr}}
\newcommand{\Jtt}{J_{\vt\vt}}
\newcommand{\Jrt}{J_{\vr\vt}}
\newcommand{\Jtr}{J_{\vt\vr}}
\newcommand{\Jvx}{J_{v_x v_x}}
\newcommand{\Jvy}{J_{v_y v_y}}
\newcommand{\Jvxvy}{J_{v_x v_y}}
\newcommand{\Jvyvx}{J_{v_y v_x}}

\newcommand{\ek}{\mathbf{e}_k}